\RequirePackage{fix-cm}

\documentclass[twocolumn]{svjour3}          

\smartqed  

\usepackage{graphicx}
\usepackage{tabularx}

\newcommand{\overbar}[1]{\mkern 1.5mu\overline{\mkern-1.5mu#1\mkern-1.5mu}\mkern 1.5mu}

\usepackage[english]{babel}      
\usepackage[T1]{fontenc}		
\usepackage[utf8]{inputenc}		%
\usepackage{hyperref}           
\usepackage{graphicx}			

\usepackage{amsmath}    
\usepackage{inconsolata}        
\usepackage[dvipsnames]{xcolor}      
\usepackage{breakcites}         
\usepackage[english,     
            linesnumbered,  
            algosection,    
            algoruled,      
           ]{algorithm2e}        
\usepackage{enumitem}       
\usepackage{amssymb}        
\usepackage{multirow}       
\usepackage{lipsum}         
\usepackage{listings}       
\usepackage[framemethod=TikZ]{mdframed} 
\usepackage{times,newtxmath}       
\usepackage[cal=cm,scr=boondoxo,bb=ams]{mathalfa} 
\usepackage{etoolbox}

\SetCommentSty{mycommentsfont}

  {\color{red}}%
  {}

\mdfsetup{
    middlelinecolor=blue!30,
    middlelinewidth=2pt,
    backgroundcolor=cyan!5,
    roundcorner=20pt,
    nobreak=true
}

\makeatletter
\hypersetup{
	pdftitle={\@title},
	pdfauthor={Matheus H. J. Saldanha and Adriano K. Suzuki},
	colorlinks=true,       		
	linkcolor=blue,          	
	citecolor=blue,        	    
	filecolor=magenta,      	
	urlcolor=blue,
	bookmarksdepth=4,
}
\makeatother

\usepackage{booktabs}
\usepackage{accents}

\renewcommand{\mid}{\,|\,}

\allowdisplaybreaks

\journalname{Statistics and Computing}

\usepackage{comment}

\begin{document}

\renewcommand{\ttdefault}{zi4}

\title{Methods to Deal with Unknown Populational Minima \\ during Parameter Inference
}


\author{Matheus Henrique Junqueira Saldanha \and Adriano Kamimura Suzuki}


\institute{M.H.J. Saldanha \at
              Institute of Mathematics and Computer Sciences \\
              University of São Paulo \\
              \texttt{\href{mailto:mhjsaldanha@gmail.com}{mhjsaldanha@gmail.com}} \\
              ORCID: 0000-0001-7701-5583
           \and
           A.K. Suzuki \at
              Institute of Mathematics and Computer Sciences \\
              University of São Paulo \\
              \texttt{\href{mailto:suzuki@icmc.usp.br}{suzuki@icmc.usp.br}} \\
              ORCID: 0000-0002-4256-4694
}

\date{Uploaded in October 16, 2020}

\maketitle

\begin{abstract}
There is a myriad of phenomena that are better modelled with semi-infinite distribution families, many of which are studied in survival analysis. When performing inference, lack of knowledge of the populational minimum becomes a problem, which can be dealt with by making a good guess thereof, or by handcrafting a grid of initial parameters that will be useful for that particular problem. These solutions are fine when analyzing a single set of samples, but it becomes unfeasible when there are multiple datasets and a case-by-case analysis would be too time consuming. In this paper we propose methods to deal with the populational minimum in algorithmic, efficient and/or simple ways. Six methods are presented and analyzed, two of which have full theoretical support, but lack simplicity. The other four are simple and have some theoretical grounds in non-parametric results such as the law of iterated logarithm, and they exhibited very good results when it comes to maximizing likelihood and being able to recycle the grid of initial parameters among the datasets. With our results, we hope to ease the inference process for practitioners, and expect that these methods will eventually be included in software packages themselves.

\end{abstract}

\section{Introduction}
\label{sec:introduction}

When performing inference on problems involving random variables with semi-infinite support, problems arise when the range of the experimental data is located far away from the origin. In this paper we analyze methods to deal with such a problem in an algorithmic, efficient, yet simple way, which does not require a case-by-case analysis to perform inference.

The aforementioned scenario can happen in various ca\-s\-es. In survival analysis, for example, the data is always supported on a semi-infinite interval, and although distributions supported on $[0, \infty)$ are the most used, there is rarely sufficient evidence that the populational minimum is indeed $0$ \cite{lawless2003statistical}. This is a reasonable assumption when the data has small location and comparatively large scale. When it has a large location and small scale, it might still be a convenient assumption for performing inference, especially if only simple location-scale or log-location-scale distributions are considered (e.g., lognormal and Weibull distributions) \cite{lawless2003statistical}. If none of these apply, the assumption leads to bad results, biased conclusions and increased difficulty in defining a reasonable initial grid of parameters for inference, as will be discussed later.

As an example, the time between failures in a supply chain might follow a Weibull with shape $\beta = 10$ and scale $\lambda = 80$, in which case there is a close to zero probability of observing a sample minimum lower than $20$ in a sample of size $100$.\footnote{$1 - (1 - F(20))^{100}$ yielding $0.0095\%$ probability, with $F(\cdot)$ being the Weibull cumulative distribution function.} Another example would be the time of a flight from Tokyo to Toronto, which clearly has a certain positive minimum value given by the limitations of airplane speed in the present age. These examples illustrate two cases that must be distinguished: one is when the underlying random variable has a long left tail; the other, when its support is $[a, \infty)$ for some unknown $a > 0$.

\begin{figure*}[t]
    \centering
    \includegraphics[width=0.8\textwidth]{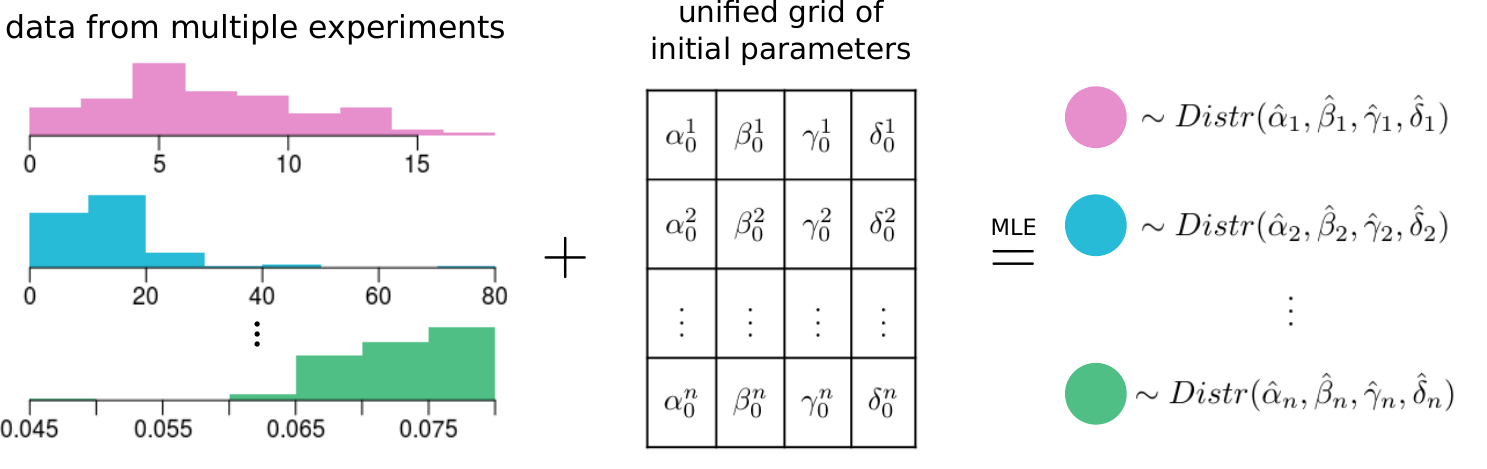}
    \caption{Main scenario to which we aim to contribute to. The experimenter has collected data from a number of different phenomena whose underlying probability distribution is believed to belong to a certain family $\mathscr{D}(\alpha, \beta, \gamma, \delta)$. We then would like to infer $\hat{\alpha}, \hat{\beta}, \hat{\gamma}, \hat{\delta}$ for each experiment. Usually, due to the variety of shapes and scales of the phenomena, one would have to define a grid of initial parameters for each phenomenon. We argue here that using our results one can define a single grid to perform inference for all the phenomena.}
    \label{fig:diagram-main-scenario}
\end{figure*}

In both these cases, it is most common to try to infer the underlying distributions using positively supported models (e.g., gamma, lognormal, Weibull), maybe after subtracting the experimental data by a certain value $c$ that the statistician believes is the theoretical minimum of the underlying distribution. In either scenario, if the underlying distribution has a long left tail, then optimizing the likelihood becomes a problem, as it can be difficult if good initial conditions are not given. Of course, simple models can be given initial conditions based on method of moments, but the same cannot be said about more complex models such as generalized versions of gamma and Weibull~\cite{stacy1962generalization,mudholkar1993exponentiated}, nested models (e.g., Kumaraswamy- and logistic-generalized distributions~\cite{cordeiro2011new,torabi2014logistic}),
mixture models~\cite{lindsay1995mixture},
etc.

This is a dangerous situation when trying to seek the model that best fits the experimental data, as one often relies on the maximized likelihood or some information criteria for model selection \cite{anderson2004model};
because of that, it is a must that the maximized likelihood be indeed the maximum, which can be made impossible if good initial conditions are not given. This could in turn lead to biased conclusions in favor of the simpler models, which are less prone to optimization issues due to bad initial conditions. In other words, it effectively renders usage of complex models useless. We therefore argue that, in these problematic cases, the sample should be modified in some way in order to simplify the determination of good initial conditions. In this paper we propose, analyze and experiment with multiple methods.

An attempt was made to imbue these methods with a reasonable amount of theoretical support, using results such as the law of iterated logarithm or asymptotic properties of the maximum likelihood estimator (MLE). Nonetheless, we allowed some room for informal reasoning, in the style of how $25$ (or $30$) is accepted as a sufficient sample size for the central limit theorem to \textit{usually} hold \cite{walpole1993probability}, or how the whiskers of a box-plot \textit{usually} serve as a good detector of outliers \cite{hoaglin2003john}. This tolerance allowed us to devise semiparametric approaches that will \textit{usually} work, as was confirmed experimentally. They are here assessed under the following objectives:
\begin{itemize}
    \item make it easier to determine a set or grid of initial values;
    \item obtain higher overall maximized likelihood over multiple models and datasets;
    \item make it possible to recycle the same grid of initial parameters for performing inference over multiple datasets, as illustrated in Fig.~\ref{fig:diagram-main-scenario}; and
    \item not incur higher computational time required for inference.
\end{itemize}

There does not seem to exist approaches, for the problem outlined above, that manage to comply with these objectives. Any parametric quantile estimator can be used for such purposes, but all estimators found either require assumptions in the underlying distribution of the random variable, such as in \cite{valk2018high,drees2003extreme,gardes2018tail,chavez2018extreme}, or they are computationally expensive, often due to usage of resampling techniques (e.g., \cite{dong2017quantile,kala2019quantile,minasny2006conditioned,liu2012convergence}).
A more comprehensive overview of related work is deferred to Sec.~\ref{sec:related-work}. The next section formalizes the problem and discuss some of its mathematical nuances and properties.
Sec.~\ref{sec:proposed-methods} presents the proposed methods to modify a sample and facilitate inference. Experiments are presented in Sec.~\ref{sec:4-experiments}, and Sec.~\ref{sec:conclusion} offers some concluding remarks.

This paper will use $\mathscr{m}$ to denote the populational minimum, $\overline{\mathscr{m}}$ the sample minimum, $f_X$ the probability density function (pdf) of random variable $X$, $F_X$ its cumulative density function (cdf), $F_n$ the empirical cdf of a sample of size $n$, $x_q$ the $q$-quantile of $X$, $\Omega_X$ the parameter space of $X$, and $\mathcal{L}_{f_X}(\theta)$ the likelihood calculated using density $f_X( \cdot \mid \theta)$. $\hat{c}$ will be an estimate yielded by some of the proposed methods, and represents that the support of the underlying distribution should be faced as being $[\hat{c}, \infty)$, or equivalently, $\hat{c}$ should be subtracted from the sample.


\section{Problem Formalization}
\label{sec:problem-formalization}

\begin{figure}[htb]
    \centering
    \includegraphics[width=0.9\linewidth]{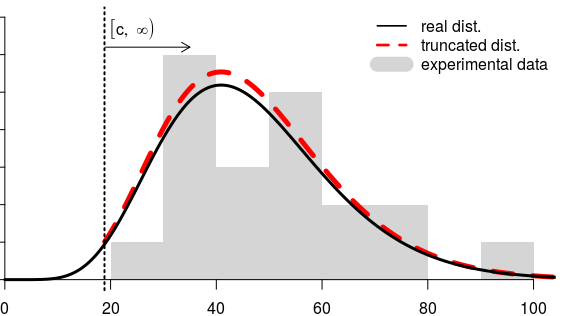}
    \caption{Example of the first scenario analyzed. The underlying phenomenon is represented by a variable $X$ whose distribution is shown as the solid line. From such a distribution we take a sample (light grey histogram), then choose $c$ using methods to be discussed later, and fit a truncated distribution over $Y \approx X - c$ (dashed line).}
    \label{fig:formulation-fig1}
\end{figure}

First let $X$ be a random variable that follows a certain probability model with support $[0, \infty)$,\footnote{Note, however, that the discussion presented here also applies to supports of type $[c, \infty)$ and $(-\infty, c]$, $c \in \mathbb{R}$.} and a sample $x_1, \dots,x_n$ taken from $X$. Consider the case where the experimental minimum is relatively high as illustrated in Fig.~\ref{fig:formulation-fig1}. By support we mean the set on which the probability density is not zero, apart maybe from a subset of measure zero; hereafter, we consider all probability functions to be defined on the whole real line. In an attempt to reduce the space of initial conditions to explore, we model such variable as $X \approx c + Y$ with $Y \in [0, \infty)$ and $c \in \mathbb{R}_+$. Note that if this model was true, the support of $X$ would be $[c, \infty)$, which violates the initial assumptions. However, it seems reasonable to believe that if $P(X < c)$ is very low, then the loss incurred by such approximation would be negligible.

The approximation here consists of considering that the range of possible outcomes of $X$ begin at a certain $c$ that is not the true one.
We then would like to model the data under such a consideration; that is, find a model for $Y$.
If we have knowledge about the distribution family of $X$ and that its support begins at zero, then a good fit (asymptotically) would be achieved by selecting the distribution of $Y$ as being a truncated version of the distribution of $X$ (see Fig.~\ref{fig:formulation-fig1}), given by
\begin{equation*}
    f_{Y}(y \mid \theta) = \frac{f_X(y + c \mid \theta)}{1 - F_X(c \mid \theta)}, \; y \in [0, \infty),
\end{equation*}
where $f_Y,f_X$ are densities, $F_X$ is a cdf, $\theta$ is a parameter vector and $c$ is given.
Notice that $f_Y(y \mid \theta) = \lambda f_X(y + c \mid \theta)$ for a constant $\lambda = 1 / (1 - F_X(c \mid \theta))$. Because of that, the likelihood over a sample $y_1, \dots, y_n$ is
\begin{align*}
    \mathcal{L}_{f_Y}(\theta \mid y_1, \dots, y_n) &= \prod_{i = 1}^n f_Y(y_i \mid \theta) \\
    &= \prod_{i = 1}^n \lambda f_X(y_i + c \mid \theta) \\
    &= \prod_{i = 1}^n \lambda f_X(x_i \mid \theta) = \lambda^n \mathcal{L}_{f_X}(\theta),
\end{align*}
so that any $\theta$ maximizing the likelihood function for $f_X$ will also maximize $\mathcal{L}_{f_Y}$ on its truncated version $f_Y$. This happens regardless of $c$, so if we allowed $c$ to also be optimized (it is one of our proposals), then it would be chosen to maximize $\lambda = 1 / (1 - F_X(c \mid \theta))$; from the monotonicity of $F_X$ we see that maximization happens when $c$ approaches the sample minimum $\overbar{\mathscr{m}}$. Clearly we cannot have $c > \overbar{\mathscr{m}}$ because in such a case at least one of the $y_i$ would be negative, thus making $f_Y$ and the likelihood $\mathcal{L}_{f_Y}$ be zero.

\newcommand{\of}[1]{\left(#1\right)}
\newcommand{\off}[1]{\left[#1\right]}
\newcommand{\offf}[1]{\left\{#1\right\}}

The fact that $c$ has no influence in the best parameter $\theta$ found by MLE is actually a problem here. Although truncation allows us to shift the support origin, it does not help with the original objective of making the space of initial parameters easier to design, since the good initial conditions are the same as for the original random variable $X$.


Since truncated models do not help here, we turn back to the original problem of finding a model for $X$ by modelling just $Y$. Recall that this means the support of $X$ is approximated as $[c, \infty)$, with $c$ being either fixed or given as a parameter of the distribution family.
Thus, consider that $X$ follows a certain distribution parametrized by some $\theta_1$ in the parameter space $\Omega_X$, whereas the candidate distribution family that we use is parametrized by $(\theta_2, c) \in \Omega_Y$ ($c$ can be fixed or not). We must then find the ``best'' $(\theta_2, c)$ in the parameter space. Let $x_1, \dots, x_n$ be a sample from the real distribution, then the average log likelihood of $(\theta_2, c)$ can be expressed as (recall $c < \overbar{\mathscr{m}}$):
\begin{equation*}
    \frac{1}{n} \sum_{i = 1}^n \log f_Y(x_i \mid \theta_2, c),
\end{equation*}
and as $n \rightarrow \infty$ we have, by the law of large numbers, its expected value:
\begin{equation} \label{eq:expected-likelihood}
    \int_0^\infty \log f_Y(x \mid \theta_2, c) \,d F_X(x \mid \theta_1),
\end{equation}
which we would like to maximize. With this we are seeking the model that obtains the highest expected likelihood over data generated by the real underlying distribution.

\begin{figure}[htb]
    \centering
    \includegraphics[width=\linewidth]{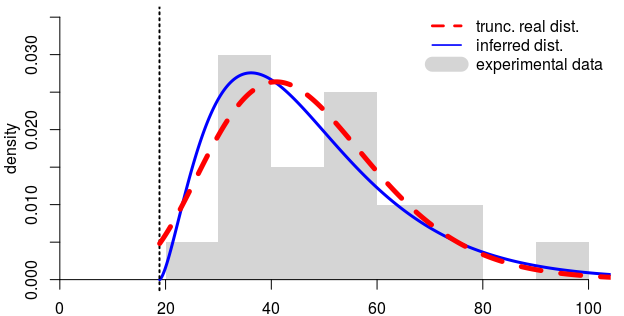}
    \caption{The ideal distribution would be the truncated version of the real underlying distribution (dashed line). However, if we exclude truncated distributions as argued in the text, we end up with a suboptimal solution (solid line), obtained here by maximizing the average likelihood shown in Eq.~(\ref{eq:expected-likelihood}).}
    \label{fig:formulation-fig2}
\end{figure}

\begin{figure}[htb]
    \centering
    \includegraphics[width=\linewidth]{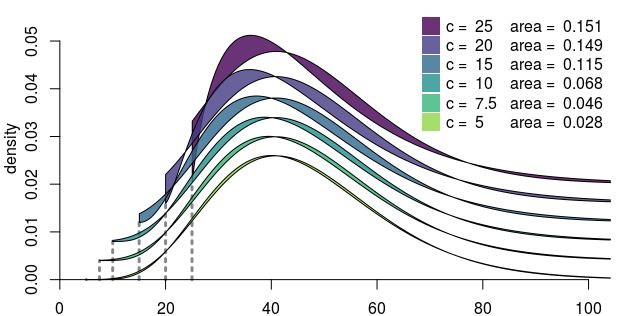}
    \caption{Considering a certain gamma distribution $\mathscr{D}$ with a long left tail, this figure shows the best gamma approximations to the ideal truncated version of $\mathscr{D}$, when performing inference on support $[c, \infty)$. Each shaded area shows the region between two curves: i) the ideal truncated distribution, and ii) the gamma distribution obtained by MLE. The curves have been displaced on the y-axis for better visualization.}
    \label{fig:areas-between-functions}
\end{figure}

Let us now consider the case where the distribution of $Y$ is inferred from the same distribution family of $X$. In this case, the optimal solution (truncated version of $X$) is usually not included in the inference search space.\footnote{Memoryless distributions are one example where it is included, and the only one that matters here. Mixtures can also be handcrafted for that to happen, but would require knowledge of the underlying distribution.} Instead, the resulting distribution will be an approximation of this optimal solution, as shown in Fig.~\ref{fig:formulation-fig2}. The area between the curves is illustrative of the difference between their cumulative probabilities, so it can be used to have an idea of how much they differ. We had constrained $c$ to be lower than the sample minimum $\overbar{\mathscr{m}}$; for small samples, this leaves a large range over which $c$ could lie. Fig.~\ref{fig:areas-between-functions} shows what happens when we perform inference for different choices of $c$. High values, nearer the sample minimum, will result in more disparate distributions than the ideal one, the truncated version of $X$. On the other hand, low values that are nearer to the origin of the original variable $X$ (lower values would also work) tend to yield a distribution more similar to the ideal. We consequently face a tradeoff as high values of $c$ is what allows recycling a grid of initial values for various phenomena.

The above discussion has so far considered that the statistician knows that the underlying random variable is supported on $[0, \infty)$. However, it is often the case that this is not known with sufficient certainty. In fact, for distributions with long left tails, which are the main object of study here, we probably will not observe any values near zero, even if the underlying distribution is indeed supported on $x \ge 0$.
This calls for methods to deal with such situations.

\section{Proposed Methods for Performing the Inference Procedure}
\label{sec:proposed-methods}

Due to the aforementioned hindrance in determining whether the underlying phenomenon is supported on $x \ge 0$, we argue that all semi-infinite random variables must be considered as belonging to an unknown interval $[\mathscr{m}, \infty)$ until proven otherwise. One could try to estimate the populational minimum, thus obtaining a support of $[\hat{\mathscr{m}}, \infty)$; however, due to what was discussed in the previous section, we seek a support $[\hat{c}, \infty)$ where we require only that $\hat{c}$ be a low quantile of the underlying distribution. As such, any value of $\hat{c}$ above the sample minimum $\overbar{\mathscr{m}}$ does not make sense. In this scenario, given a sample $x_1, \dots, x_n$ we would like to find the underlying distribution within a parametrized family supported on $[\hat{c}, \infty)$.

In order to ease the determination of initial parameters for the subsequent inference process, $\hat{c}$ is to be chosen regardless of the real value of $\mathscr{m}$, merely aiming for having $P(X < \hat{c})$ be low enough and $\hat{c}$ be as near the sample minimum as possible, due to arguments given in Sec.~\ref{sec:problem-formalization} (and seen in Fig.~\ref{fig:areas-between-functions}). Our choices are then to either to estimate $\hat{c}$ and then perform inference over $Y = X - \hat{c}$ using a family $\mathscr{D}(\theta)$, or to find $\hat{c}$ by adding a location parameter to such family, which then becomes $\mathscr{D}(\theta, c)$. We analyze both possibilities, and in the end propose a third alternative that deviates slightly from the usual procedure of classical inference. We remind that the objective is maximize likelihood, improve ease of use (i.e., make it easier to define an initial grid of parameters), and minimize computational cost.

\medskip
\textbf{I) Inferring the Location Parameter.} Let $X$ represent the underlying phenomenon with support $[\mathscr{m}, \infty)$. We want to model it using a family $\mathscr{D}(\theta)$ of $\mathbb{R}_+$ supported distributions, though shifted to $[c, \infty)$. That is, we actually model $Y$ such that:
\begin{align*}
    f_Y(y \mid \theta, c) = \begin{cases}
        f_X(y - c \mid \theta),\;&\text{ if } y \ge c \\
        0,\;&\text{otherwise},
    \end{cases}
\end{align*}
in which case, we say $Y \sim \mathscr{D}(\theta, c)$ with $c$ constrained to lie in the interval $[0, \overbar{\mathscr{m}})$, or to $[-\infty,\overbar{\mathscr{m}})$ if the experimenter deems reasonable. With this, MLE can then be performed to find $\hat{\theta}$ and $\hat{c}$. Of course, the optimizer will probably ask for an initial value of $c$, which can be done by means of the four estimators proposed in item II. The following code illustrates the inference process, using as initial value $\overbar{\mathscr{m}} - \hat{\sigma} / n$ (explained later):

\begin{lstlisting}[
  emph={min,sd,rgamma,dgamma,optim,sum,c,log},emphstyle=\it\tt\color{blue},
  emph={[2]function},emphstyle={[2]\bf\tt\color{ForestGreen}}
]
N     = 20;
data  = rgamma(n=N, shape=2.3, scale=2);
cinit = min(data) - sd(data)/N;
likelihood = function(p) - sum(log(
              dgamma(data - p[3],
               shape=p[1], scale=p[2])));
result = optim(
  par=c(1, 1, cinit), fn=likelihood);
\end{lstlisting}



\medskip
\textbf{II) Estimating the Location Parameter.} \sloppy Since the lower bound of the desired support $[c, \infty)$ is strongly related to the low quantiles of the population, it makes sense to use sample information to estimate it. With this estimate we can then perform inference using a positively supported distribution as usual (on $[0, \infty)$) after subtracting $\hat{c}$ from the sample (recall that $Y \approx X - c$). Taking the sample minimum to estimate it, besides being very biased, also frequently results in the likelihood becoming constant, rendering optimization by MLE impossible. To see this, note that after subtracting the estimate from the sample $x_1, \dots, x_n$, the smallest one ends up being $x_j - \hat{c} = \overline{\mathscr{m}} - \overline{\mathscr{m}} = 0$; the problem here is that many distributions yield problematic values for $f(0 \mid \theta)$ (i.e., $0$ or $\infty$) for a large range of their parameters, which when plugged in the log-likelihood function, makes it go to $-\infty$ if $f(0 \mid \theta) = 0$, or to $\infty$ if $f(0 \mid \theta) = \infty$. Considering the gamma distribution, for illustration, we have $f(0 \mid \theta) = 0$ when the shape parameter is $\alpha > 1$, and $f(0 \mid \theta) = \infty$ when it is $\alpha < 1$. 

Shifting that estimate slightly to the left is thus needed, maybe by multiplying it by some factor. But what should this factor be? In our experience, deciding this automatically to various datasets with different shapes and scales happened to be quite difficult. For example, taking $\hat{c}$ to be $0.95 \overbar{\mathscr{m}}$ worked for datasets with smaller values, but not for larger ones where it was shifted too far from the sample minimum. In order to find better alternatives, we first improve the above $0.95 \overbar{\mathscr{m}}$ estimate, and then later rely on order statistics.

The sample minimum $\overbar{\mathscr{m}}$ has a known cumulative distribution:
\begin{align} \label{eq:minimum-cdf}
    F_{\overbar{\mathscr{m}}}(x \mid \theta) = 1 - [ 1 - F_X(x \mid \theta) ]^n,
\end{align}
for a sample of size $n$ of a variable with cdf $F_X(x \mid \theta)$. For $n = 1$ we have the same distribution as $X$, and for $n \rightarrow \infty$ it converges in distribution to the populational minimum~\cite{haan2007extreme}, as we are dealing with continuous models (a finite number of discontinuities is also tolerable). Thus, the sample minimum begins with the variance of the random variable, and ends with zero variance; in the interim, the variance decreases at a certain unknown rate. This reasoning brought us to the first estimator, which is more informal than the others, but worked well in practice. Contrary to the other three, this estimator is similar to what is known as multiplicative quantile estimators~\cite{yazidi2017multiplicative}, which assumes that the statistician can be sure that the underlying random variable is positive; that is, if the populational minimum is negative, it will not work. The estimator is defined as follows:
\begin{align} \label{eq:c1}
    \hat{c}_1(x_1, \dots, x_n) = \overbar{\mathscr{m}} \cdot \left( 1 - \frac{\hat{\sigma}}{\hat{\mu} \log_k(n)} \right),
\end{align}
where $\hat{\sigma} / \hat{\mu}$ is the variation coefficient of the sample and $k$ is an arbitrary logarithm basis. The interpretation is that we are moving $\overbar{\mathscr{m}}$ towards the origin, with an intensity that is directly proportional to the data variability and inversely proportional to the sample size.

Our experience showed $k = 10$ to be quite useful. To see the implications of other choices of $k$, note that Eq.~(\ref{eq:c1}) can be rewritten, by a change of logarithm basis, as:
\begin{align*}
    \overbar{\mathscr{m}} \cdot \left( 1 - \log_{10}(k) \frac{\hat{\sigma}}{\hat{\mu} \log_{10}(n)} \right),   
\end{align*}
so there is a difference of a constant factor $\log_{10}(k)$. For illustration, $\log_{10}(e) = 0.434$, so the estimator will approach the sample minimum with about double the speed; we see that one could very much choose a value for $\log_{10}(k)$ directly, instead of choosing $k$. Besides these considerations, it is also worth noting that taking the coefficient of variation eliminates, to a certain extent, problems caused by the scale of the data, since it involves a division by the sample mean. This estimator has the advantage of simplicity, and even though it is not backed by a strong theoretical foundation, it appears to work very well in practice.

We now turn to more complex alternatives, that have more theoretical grounds. Although there are many parametric approaches for estimating quantiles, estimating low ($<0.05$) quantiles is a problem that has not yet been solved in a sufficiently general way. That is, most parametric solutions rely on assumptions about the underlying distribution or quantile functions (constraints on the derivative of the pdf, for example \cite{daouia2007nonparametric,mu2007power,alexopoulos2019sequest}). To maintain generality (and because this later proved to work well), we opt for more general semiparametric approaches, using the empirical cdf $F_n(x)$ over $n$ samples as main tool. Uniform convergence of $F_n(x)$ to $F(x \,|\, \theta)$ is given by the Glivenko-Cantelli theorem,\footnote{This, as well as all other results used hereafter, require independent and identically distributed sampling.} so for sufficiently large $n$ we have information about the probability $P(Y \le \overbar{\mathscr{m}}) \approx F_n(\overbar{\mathscr{m}}) = 1/n$ of a next sample to be lower than the current sample minimum. As this number decreases, the less we can expect the populational minimum to be lower than the actual sample minimum, meaning that we can then define a second estimator:
\begin{align} \label{eq:estim-one-by-n}
    \hat{c}_2(x_1, \dots, x_n) = \overbar{\mathscr{m}} - \frac{\hat{\sigma}}{n},
\end{align}
where we embody the hope that the deviation between populational and sample minimum be proportional to the sample standard deviation $\hat{\sigma}$ and to $F_n(\overbar{\mathscr{m}})$. Note that it is an additive estimator, which is a choice based on good experimental results and on dimensional analysis~\cite{goldberg2006fundamentals}; since $\hat{\sigma}$ and $\overbar{\mathscr{m}}$ have the same measurement unit, it makes sense to subtract them. In contrast, $\hat{c}_1$ uses the coefficient of variation, which is dimensionless and thus more suitable as a multiplicative constant.

A tighter estimate follows by noticing that the law of iterated logarithm~\cite{vapnik1998statistical} gives the rate of convergence:
\begin{align*}
    \forall x \, \lim_{n \rightarrow \infty} \; \sup_{l > n} \; \big| F(x \mid \theta) - F_l(x) \big| \le \sqrt{\frac{\ln \ln l}{2 l}}.
\end{align*}
Now using that $F_n(\overbar{\mathscr{m}} - \epsilon)$ is zero for any $\epsilon > 0$, we must have for sufficiently large $n$:
\begin{align*}
    F(\overbar{\mathscr{m}} - \epsilon \mid \theta) \le F_n(\overbar{\mathscr{m}} - \epsilon) + \sqrt{\frac{\ln \ln n}{2 n}} \longrightarrow \sqrt{\frac{\ln \ln n}{2 n}},
\end{align*}
which can then substitute the $1/n$ in Eq. (\ref{eq:estim-one-by-n}):
\begin{align} \label{eq:estimator-iter-logarithm}
    \hat{c}_3(x_1, \dots, x_n) = \overbar{\mathscr{m}} - \hat{\sigma} \cdot \sqrt{\frac{\ln \ln n}{2 n}}.
\end{align}

\begin{figure}[htb]
    \centering
    \includegraphics[width=\linewidth]{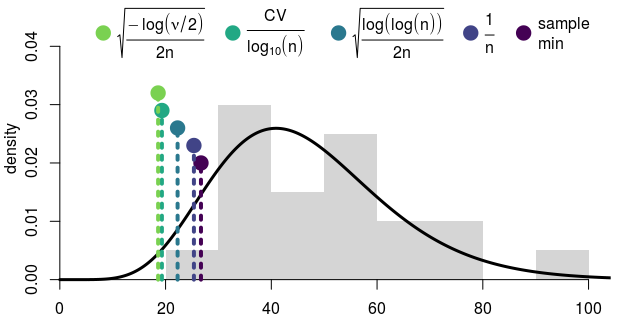}
    \caption{The low quantile estimators based on the data represented by the histogram in light gray. The data was generated from the density shown as a black line. Here we use $\nu = 0.05$.}
    \label{fig:estimators-example}
\end{figure}

The Dvoretzky-Kiefer-Wolfowitz inequality \cite{massart1990tight} can al\-so be invoked, which provides a different way to view the estimator. The inequality is:
\begin{align*}
    P( \sqrt{n} \sup_x | F_n(x) - F(x \mid \theta) | \le \lambda ) \ge 1 - 2 \exp(-2 \lambda^2),
\end{align*}
and by doing the necessary manipulations, we derive that the following will hold with probability of at least $1 - \nu$:
\begin{align*}
    \sup_x | F_n(x) - F(x \mid \theta) | \le \sqrt{\frac{-\ln(\nu / 2)}{2n}},
\end{align*}
so if we choose $\nu$ to be very low, we can expect $F(\overbar{\mathscr{m}} - \epsilon \mid \theta)$ to be lower than or equal to the right-side of the above equation. Following the same logic as previously, we define another estimator:
\begin{align} \label{eq:estimator-dvoretzky}
    \hat{c}_4(x_1, \dots, x_n) = \overbar{\mathscr{m}} - \hat{\sigma} \cdot \sqrt{\frac{-\ln(\nu / 2)}{2n}},
\end{align}
which offers a probabilistic view, instead of the previous asymptotic view given by the Law of Iterated Logarithm. Fig.~\ref{fig:estimators-example} illustrates all of these estimators.

\medskip
\textbf{III) Iterative Determination of the Location Parameter.} Inference by MLE begins with the assumption that the underlying distribution comes from a certain family. Under this assumption, we do have a lot of information about the underlying cdf and pdf. We intend to use this information to our advantage here.

The cdf of the sample minimum is given by Eq. (\ref{eq:minimum-cdf}). By inverting that equation we obtain the quantile function of the minimum:
\begin{align} \label{eq:quantile-of-minimum}
    F_{\overbar{\mathscr{m}}}^{-1}(q \mid \theta) = F_X^{-1}(1 - (1 - q)^{1/n} \mid \theta).
\end{align}
With this, the median of the sample minimum is given by $F_{\overbar{\mathscr{m}}}^{-1}(0.5 \mid \theta)$, under the assumptions that the underlying distribution resides in the specified family and has parameter $\theta$. The median can be seen as a good guess for what the sample minimum should be, and so the sample should be shifted so that the sample minimum coincides with such a guess. When performing MLE, the subtraction is done on every iteration of the optimization algorithm, right before calculating the log-likelihood. The following R code illustrates the process:
\begin{lstlisting}[
  emph={min,sd,rgamma,dgamma,qgamma,optim,sum,c,log},emphstyle=\it\tt\color{blue},
  emph={[2]function},emphstyle={[2]\bf\tt\color{ForestGreen}}
]
N = 20;
data = rgamma(n=N, shape=2.3, scale=2);
likelihood = function(p){
  q = qgamma(1 - (1 - 0.5)^(1/N),
        shape=p[1], scale=p[2]);
  -sum(log(dgamma(data - q,
             shape=p[1], scale=p[2])));
}
result = optim(
  par=c(1, 1), fn=likelihood);
\end{lstlisting}




\section{Experimental Results}
\label{sec:4-experiments}

In order to reason about what is a good way to assess the estimators, recall that we have two opposing objectives:
\begin{enumerate}
    \item[i)] $\hat{c}$ should be as near the sample minimum as possible, and
    \item[ii)]the cumulative probability $P(X < \hat{c})$ should be as low as possible.
\end{enumerate}
One could imagine that using $F^{-1}(0.01 \mid \theta)$ (or $F^{-1}(\nu \mid \theta)$ for any small $\nu$ defined by the user) as the ideal value would manage to fulfill both objectives. However, for any $\nu$ there will be a sample size $n$ that makes the sample minimum be below $F^{-1}(\nu \mid \theta)$ with high probability, and it does not make sense to take as ideal value of $c$ a number that is above the sample minimum. Thus, some adapting rule based on $n$ must be included.

\begin{figure}[htb]
    \centering
    \includegraphics[width=\linewidth]{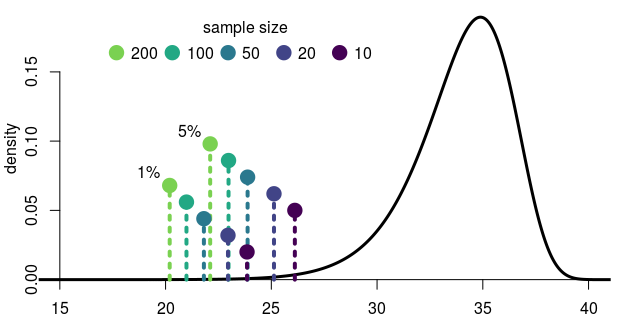}
    \caption{This illustrates the $5\%$ and $1\%$ quantiles for the sample minimum from a sample taken from the distribution shown in black. Note on the $x$-axis that the distribution has a large location.}
    \label{fig:minimum-quantiles}
\end{figure}

To take the sample size $n$ into consideration, we find it better to use the distribution of the sample minimum. Consider the quantile function for the sample minimum of a sample of size $n$, as given in Eq.~(\ref{eq:quantile-of-minimum}). Then there is a $1\%$ probability to obtain a minimum lower than $q_{0.01} = F_{\overbar{\mathscr{m}}}^{-1}(0.01 \mid \theta)$; thus, under the assumption that $\theta$ is the true parameter, this number will \textit{probably} be located to the left of the sample minimum $\overbar{\mathscr{m}}$. For $n = 1$ it coincides with $F^{-1}(0.01 \mid \theta)$, so it can be seen as analogous to using $F^{-1}(\nu \mid \theta)$ as discussed above, but which adapts to the sample size. We can also expect this $q_{0.01}$ not to be located too deep into the left tail of the distribution,
which is illustrated in Fig.~\ref{fig:minimum-quantiles}. Therefore it seems that $q_{0.01}$ fulfills both desired properties i) and ii) presented earlier, and is thus a good baseline to which to compare our estimates, and we use it hereafter. It is also advantageous for computer experiments because it does not change from one experiment to another, as it depends on the actual distribution, and not on a random sample thereof.

\medskip
\noindent \textbf{A) Tests on wine quality dataset}
\smallskip

\begin{figure}[htb]
    \centering
    \includegraphics[width=0.8\linewidth]{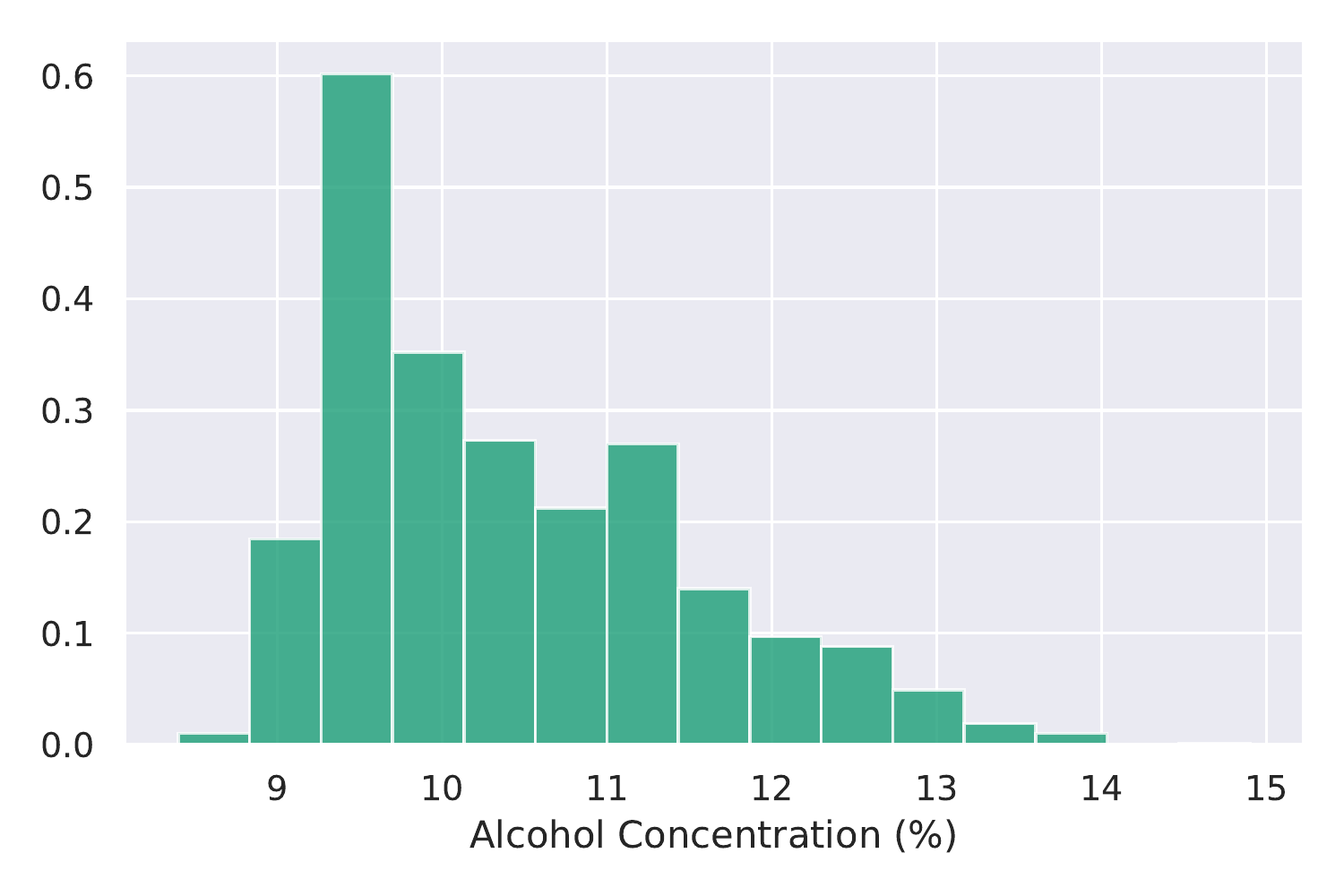}
    \caption{Histogram of the wine dataset, showing the alcohol concentration of multiple bottles of wine.}
    \label{fig:wine-histogram}
\end{figure}

\noindent The first practical scenario considers a dataset containing characteristics of $1599$ bottles of the same brand of a portuguese red wine \cite{cortez2009modeling}. In particular, we analyze their alcohol concentration, whose histogram is displayed in Fig.~\ref{fig:wine-histogram}. It has the characteristic of having a large location, and it is easy to believe that it is a positive random variable, maybe with a populational minimum that is not zero. In order to determine the underlying distribution, we perform MLE using nine distributions: gamma, Weibull, normal, truncated normal, lognormal~\cite{lawless2003statistical}, odd log-logistic generalized gamma (OLL-GG)~\cite{prataviera2017odd}, Kumaraswamy complementary Weibull geometric (Kw-CWG)~\cite{afify2017new}, generalized gamma~\cite{stacy1962generalization} and generalized Weibull~\cite{mudholkar1993exponentiated}. Our objective is to show that the proposed estimators improve likelihood and to analyze their computational cost, especially when considering complex, $5$-parameter models.

Table~\ref{tab:wine-AIC} shows the goodness-of-fit values, in this case the Akaike information criterion (AIC), obtained for each method to deal with the population minimum. Experiments for the $20$ and $100$ sample sizes were repeated 5 times, each with a different sample taken from the whole dataset.
We are not particularly interested in which distribution family achieved best fit; rather, only in what are the best fits obtained when each method of dealing with the populational minimum is used.
Therefore, Table~\ref{tab:wine-AIC} shows the $5\%$ and $90\%$ quantiles of all likelihoods obtained, as a means to show a more robust estimate of the capability of each method to help finding good fits. For completeness, we note that for sample size $n = 20$ the \texttt{iterated} method achieved the best AIC in all $5$ experiment trials; for $n = 100$, \texttt{infer c} won in $4$, and \texttt{c3} in $1$ trial.

It is clear that inconsequentially considering the populational minimum as being zero leads to worse results, as evidenced on the $90\%$ quantiles shown in Table~\ref{tab:wine-AIC}. This is precisely the case where the more complex models are rendered useless (as mentioned in Sec.~\ref{sec:introduction}), as they were the ones with which the optimization algorithm often could not converge, resulting in absurd values of the AIC. It might make sense to deem the alcohol concentration as a random variable $X > 0$ with a long left tail, but it just places too much burden on the optimization algorithm to navigate the rough parameter surface to try to model the data correctly. It leads to a lot of cases (in Table~\ref{tab:wine-AIC}, at least $10\%$) where the optimization procedure diverges, mainly because the optimal parameters for modelling data with high location and low relative variance tend to be absurd (e.g., $\alpha = 4000$ and $\beta = 1/500$ would not be surprising for a gamma), which is not generally easy to converge to, given the grid of initial values defined by the experimenter. This is a major problem for the more complex distributions and, in fact, all divergent cases came from the distributions with $3$ or more parameters.

\begin{table}[tb]
    \centering
    \renewcommand{\arraystretch}{1.2}
    \begin{tabular}{@{}rlll@{}} \toprule
        & \multicolumn{3}{c}{Sample size} \\ \cmidrule{2-4}
        Method \hspace{1em} & 20 &  100 & 1599 \\ \midrule
        no estimation & $53.7$ -- $16\mathrm{K}$ & $284.1$ -- $83\mathrm{K}$ & $4511$ -- $445\mathrm{K}$ \\
        $\hat{c}_1$ & $49.4$ -- $71.1$ & $268.0$ -- $300.6$ & $4324$ -- $4693$ \\
        $\hat{c}_2$ & $49.5$ -- $71.3$ & $268.7$ -- $301.4$ & $4325$ -- $4701$ \\
        $\hat{c}_3$ & $45.8$ -- $69.8$ & $264.7$ -- $297.7$ & $4334$ -- $4648$ \\
        $\hat{c}_4$ & $47.6$ -- $69.2$ & $265.8$ -- $\mathbf{296.6}$ & $4330$ -- $4650$ \\
        iterated & $\mathbf{-30.6}$ -- $\mathbf{65.1}$ & $243.1$ -- $310.6$ & $4354$ -- $4856$ \\
        infer $c$ & $-5.5$ -- $66.6$ & $\mathbf{222.5}$ -- $296.8$ & $\mathbf{4319}$ -- $\mathbf{4643}$ \\
        \bottomrule
    \end{tabular}
    \caption{AIC values obtained by performing MLE on the wine dataset. They show the $5\%$ and $90\%$ quantiles of the AIC obtained, considering each way to handle the populational minimum. Recall that lower values are better. Best values in each column are highlighted in bold; the notation $\mathrm{K}$ is used to denote thousands ($10^3$).}
    \label{tab:wine-AIC}
\end{table}

Table~\ref{tab:wine-AIC} shows that the proposed methods outperform the baseline, as expected. For small sample sizes, the iterated method and adding $c$ as a distribution parameter clearly overfitted the data by setting $\hat{c}$ as close as possible to the sample minimum. In this sense, the $\hat{c}_k$ estimators proved themselves to be the most ``stable'', displaying good performance for any sample size. Even $\hat{c}_2$, the crudest method, managed to keep up with the best (non-overfitted) AIC values. Inferring $c$ yielded the best results for samples of size $100$ and $1599$, though some care must be taken due to the increase in parameter count. The iterated method only displayed advantage in size $100$, so it did not prove to be a safe choice for this kind of problem. We highlight that for size $1599$ it had a relatively large variance in the AIC values -- seen in the table as the difference between the two values in its cell --, which is one downside of this method.

Table~\ref{tab:wine-time} shows the computational time taken to perform MLE when using each method, and shows that the $\hat{c}_k$ methods tend to be faster than the other proposed methods. The `no estimation' method appears to be fast, but it is because inference stops very quickly when the optimization diverges, and this method had many divergent cases. In the table, it is also notable that the iterated method is relatively slower than the other methods in small sample sizes, but it becomes quite competitive when considering the whole dataset.

\begin{table}[tb]
    \centering
    \renewcommand{\arraystretch}{1.2}
    \begin{tabular}{@{}rlll@{}} \toprule
        & \multicolumn{3}{c}{Sample size} \\ \cmidrule{2-4}
        Method \hspace{1em} & 20 &  100 & 1599 \\ \midrule
        no estimation & $\mathbf{0.40}$ & $0.51$ & $\mathbf{2.23}$ \\
        $\hat{c}_1$ & $0.43$ & $0.53$ & $2.81$ \\
        $\hat{c}_2$ & $0.42$ & $0.54$ & $2.82$ \\
        $\hat{c}_3$ & $0.41$ & $\mathbf{0.47}$ & $2.50$ \\
        $\hat{c}_4$ & $0.41$ & $0.48$ & $2.67$ \\
        iterated    & $0.65$ & $0.66$ & $2.87$ \\
        infer $c$   & $0.58$ & $0.68$ & $3.54$ \\
        \bottomrule
    \end{tabular}
    \caption{Computational time (in minutes) to perform the whole MLE process when using each method, considering the wine dataset. Since the experiment was repeated five times for sample sizes $20$ and $100$, the total time has been divided by $5$. Best values in each column are highlighted in bold. The experiments were performed in an idle machine, with a CPU Intel i7 860 2.80GHz.}
    \label{tab:wine-time}
\end{table}

\medskip
\noindent \textbf{B) Tests on execution times dataset}
\smallskip

\noindent In the following, results are presented concerning the main scenario to which the proposed estimators were designed to contribute. Consider a study of the probability distribution of the execution time of a certain deterministic mathematical computer program, such as calculation of the Mandelbrot set~\cite{alligood1996chaos}. For this, the program is executed a thousand times in $n$ different machines $M_1, \dots, M_n$, generating $n$ datasets. The experimenter wants to determine whether there is a probability distribution that best models \textit{all} of these datasets, so they perform MLE in each of these datasets using multiple distributions. The variety of machines is large due to the number of different vendors and versions of CPUs, motherboards and RAM memory, so $n$ is large (this problem is analyzed in more detail in \cite{saldanha2020probabilistic,saldanha2020execution}). Clearly, this is a time-consuming process. In our experience, it becomes worse because it is significantly difficult to define a initial grid of parameters that will lead MLE to converge nicely for all datasets, due to the large variety in locations and variances of the samples. Moreover, the execution time of a program is clearly a variable of type $X > c$, for some populational minimum $c > 0$, since there is a physical limitation on the smallest time that the program can execute in any machine, and from the perspective of inference this is another hindrance to deal with.

In these circumstances, one option is to analyze each dataset individually, defining initial grids of parameters for each distribution family, and making a guess for the populational minimum $\hat{c}$ based on the dataset histogram, for example. Fortunately, if our proposed estimators are used in this situation, not only it allows for algorithmically determining a good guess $\hat{c}$ for the populational minimum, but in our experience it also helps devising a single initial grid of parameters that will work for all datasets. Roughly, when defining an initial grid of parameters one has to anticipate the location, scale and shape of the data; the logic here is that once the dataset is subtracted from populational minimum, one may ignore the location and focus on the other two aspects only.

While it is difficult to convey, through numbers, the improvement in experience and productivity, we try to show one result that somewhat corroborate these assertions. We performed the experiment described above, with $37$ different datasets (i.e., $37$ machines) and using $9$ different distribution families: gamma, Weibull, normal, truncated normal, lognormal, the aforementioned OLL-GG and Kw-CWG, generalized gamma and generalized Weibull. The likelihoods obtained cannot be directly compared due to large differences in the scale of the datasets, so some transformation of the likelihood is necessary. For each dataset, we consider the best log-likelihood (out of $9$, one per distribution family) obtained by each inference method; furthermore, for each dataset, one of the methods achieved the best log-likelihood, so the performance of each method can be measured as the term $\hat{l} - \hat{l}_b$ where $\hat{l}$ is the best log-likelihood of the method and $\hat{l}_b$ is the best likelihood obtained among all methods. The difference of log-likelihoods is directly related to the ratio of likelihoods, which in turn has known asymptotic properties \cite{severini2000likelihood} so this transformation has theoretical ground.

\begin{figure*}
    \centering
    \includegraphics[width=0.75\linewidth]{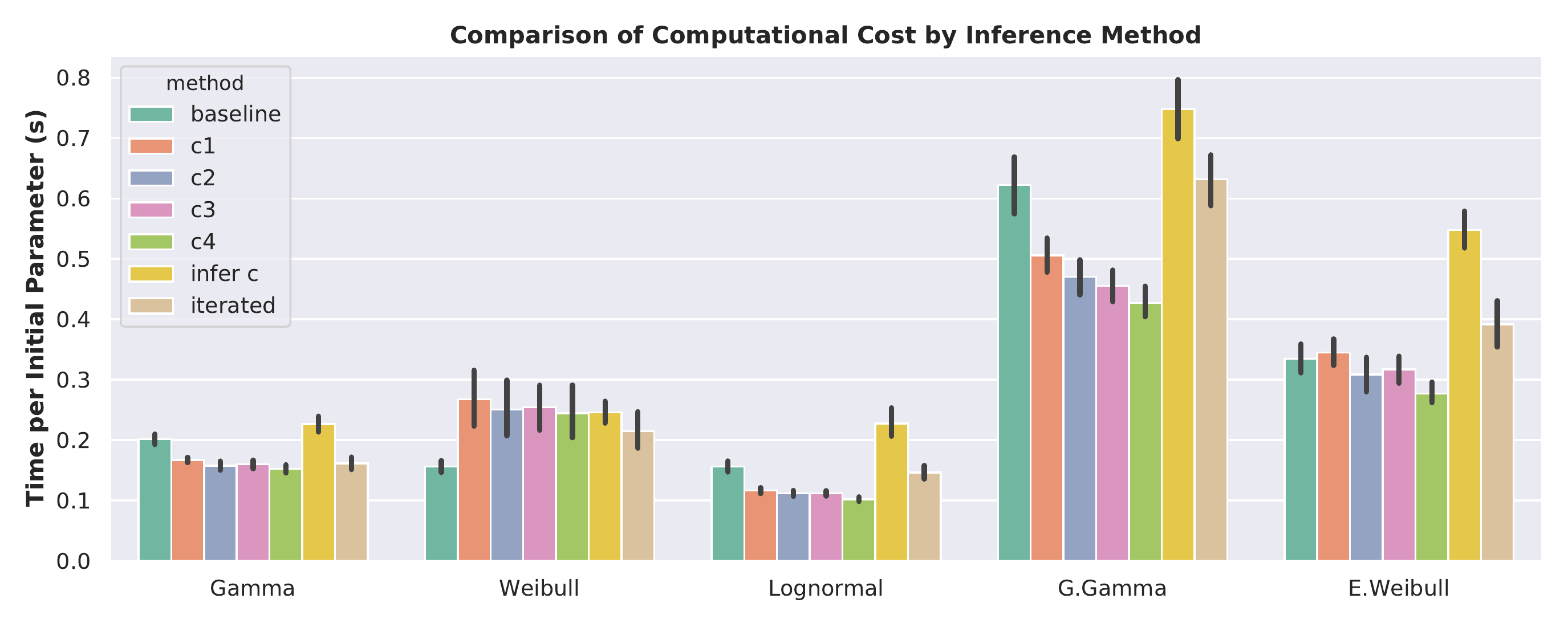}
    \caption{For five distribution families considered, the figure shows the computational cost per MLE optimization procedure, for each method of modifying the sample, considering the dataset of execution times of programs. This considers the average time taken for one initial vector of parameters; naturally, more complex distributions involve a larger grid of initial parameters, and thus require more executions of the optimization procedure, consequently leading to a more lengthier process.}
    \label{fig:computational-cost-stochastic-scheduling}
\end{figure*}

\begin{figure}[htb]
    \centering
    \includegraphics[width=0.9\linewidth]{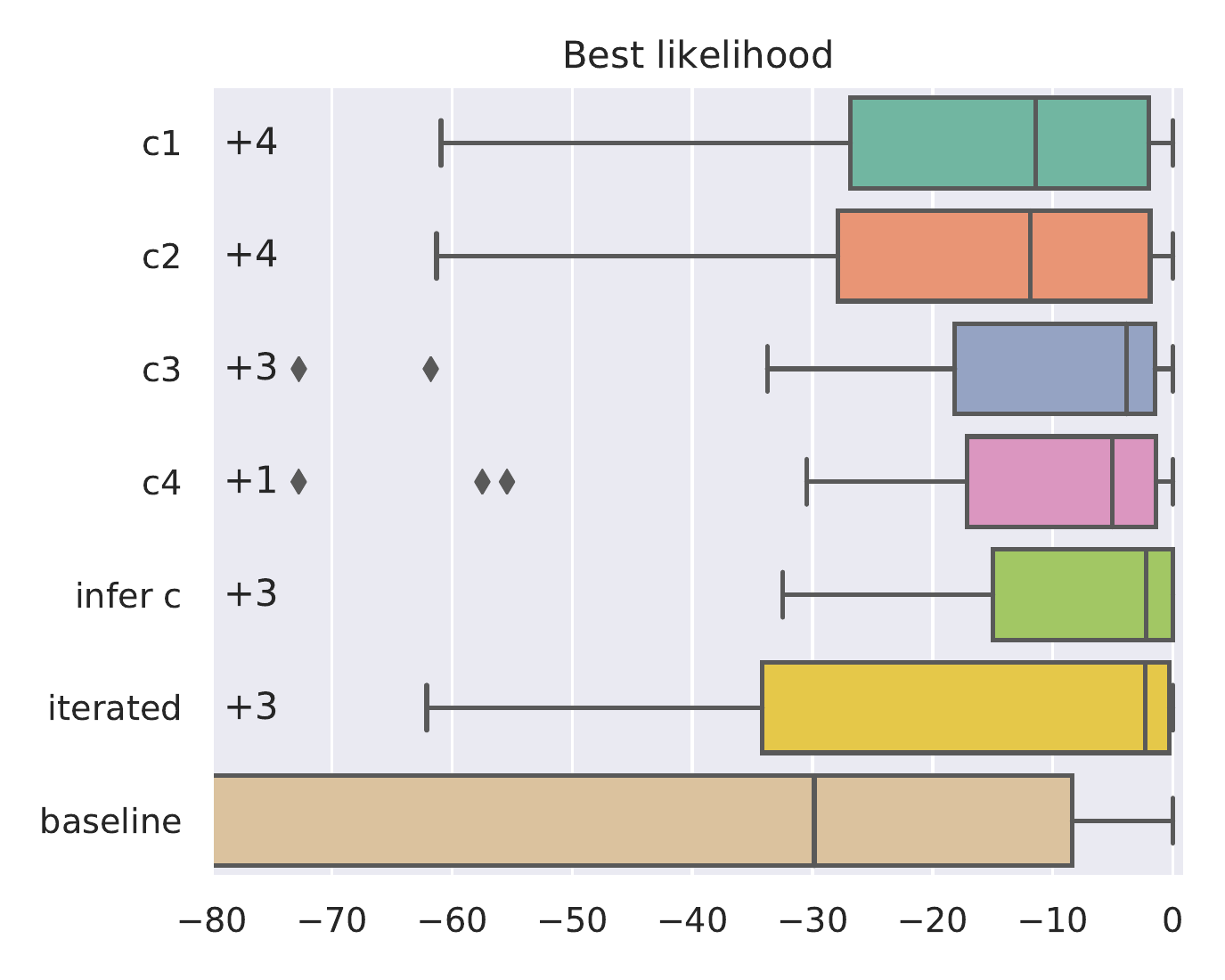}
    \caption{Log-likelihood differences showing the performance of each method on each of the $37$ datasets. A value of $0$ shows that the method achieved the best performance in one of the datasets. Numbers on the left border are the number of outliers outside the area shown by the plot.}
    \label{fig:log-likelihood-differences}
\end{figure}

These log-likelihood differences are shown in Fig.~\ref{fig:log-likelihood-differences}, in which a value of $0$ shows that the method in question was the best performing in one of the $37$ datasets; negative values show how far from the best method it was. It is noticeable that the proposed methods outperformed the baseline method considerably, which is an indicative of the benefits brought by using the proposed estimators.
Note that the poor performance of the baseline has not been caused by a poor choice of the initial grid of parameters. The same grid was used for all models, and it was designed to cover distributions of various locations, scales and shapes, as we would do if using just the baseline model.
We consider our efforts to have been successful, since in most cases the \texttt{baseline} method indeed does converge to some set of parameters, though often suboptimal ones. The reader can check the initial grid of parameters in one of our code repositories.\footnote{See the file \texttt{wine-quality/allModels.r} in \url{https://github.com/matheushjs/dealing-with-popmin}.}

We also highlight that the median of $\hat{c}_3$, $\hat{c}_4$, \texttt{infer c} and \texttt{iterated} methods are all very close to zero, the best performance, which means that these were the best methods for this scenario, setting \texttt{iterated} aside due to its increased variance. Our assessment of the boxplot outliers is as follows: most cases occurred in datasets that contained an outlier, where the extra flexibility of the \texttt{infer c} method allowed it to achieve better results than the $\hat{c}_k$ estimators; the outliers for the \texttt{infer c} have happened in datasets that displayed two modes, a small mode located to the left of the main mode, and here the other methods achieved superior likelihood values. Overall, however, the visual inspection of the histograms did not indicate disparities as large as the likelihood values make it seem, that is, if we could ignore a few samples of each dataset, most outliers in Fig.~\ref{fig:log-likelihood-differences} would not occur.

For another point of view, Fig.~\ref{fig:computational-cost-stochastic-scheduling} shows the average time taken per MLE optimization for each method, which shows that \texttt{infer c} is slower, as expected. It adds one extra degree of freedom, which places more burden on the optimization algorithm. 
For similar reasons, the \texttt{iterated} method is slightly slower than the other methods. The \texttt{baseline} method is slower for some distributions, and faster for others; the reasons for this depends mostly on how often the MLE inference diverged for these distributions. When considering the average only of the non-diverging cases, the \texttt{baseline} shows a similar computational cost than the $\hat{c}_k$ methods. The $\hat{c}_k$ proved to be quite fast; $\hat{c}_4$ displayed a small lead over the others, but it is safer to attribute this to particularities in their implementations. Also, we highlight that the difference between the baseline and the proposed methods is larger in practice than what is shown in Fig.~\ref{fig:computational-cost-stochastic-scheduling}. This is because the grid of initial parameters for the \texttt{baseline} method will have to be larger in order for it to work well with multiple datasets, whereas the proposed methods were designed to make such grid smaller. Thus, if the bars in the figure are multiplied by the size of the grid (i.e., the total number of calls that would be made to the optimization procedure), there will be an astounding superiority of our proposed methods relative to the \texttt{baseline}.

\begin{figure*}
    \centering
    \includegraphics[width=0.4\textwidth]{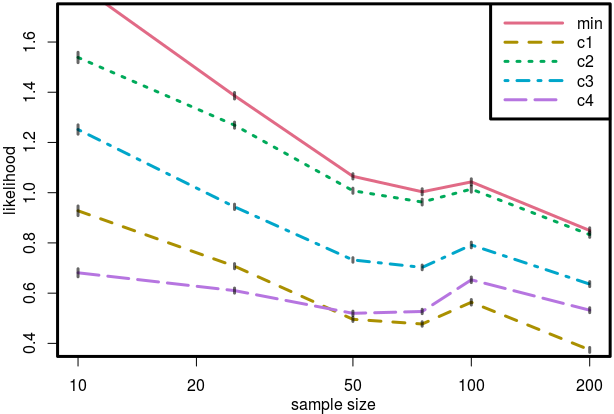}
    \hspace{2em}
    \includegraphics[width=0.43\textwidth]{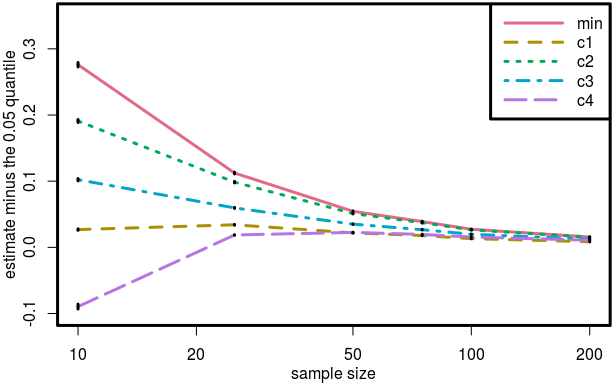}
    \caption{Performance of the estimators $\hat{c}_1$, $\hat{c}_2$, $\hat{c}_3$ and $\hat{c}_4$ on an $\mathit{Exp(1/3)}$, for samples of different sizes. (left) The log-likelihood obtained by fitting an exponential to the data subtracted from each estimator. Each point is subtracted from the log-likelihood obtained by the $\mathit{Exp(1/3)}$ on the data sample. (right) Signed distance from each estimate to the $5\%$ quantile of the sample minimum distribution. Negative values show an estimate that was lower than the quantile. Note that the $x$-axis is on log-scale, and error bars show a $99\%$ confidence interval.}
    \label{fig:test-exponential}
\end{figure*}

\begin{figure}[htb]
    \centering
    \includegraphics[width=0.9\linewidth]{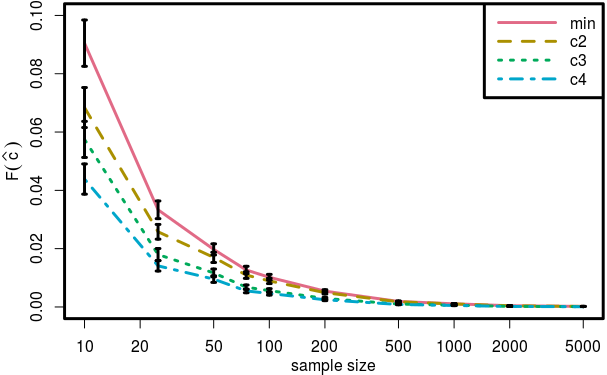}
    \caption{Probability of taking a sample lower than the estimates (i.e., $F(\hat{c})$) for a $\mathit{Cauchy}(0,1)$. The $x$-axis is in log-scale, and the errors bars show $0.1 \hat{\sigma}$ (one tenth of the sample standard deviation) in each side, of 200 replications of the experiment.}
    \label{fig:test-cauchy}
\end{figure}

\medskip
\noindent \textbf{C) Two worst-case synthetic scenarios}
\smallskip

\noindent We now assess the performance of the estimators ($\hat{c}_1$, $\hat{c}_2$, $\hat{c}_3$, $\hat{c}_4$) in two worst-case scenarios. The first is the exponential distribution, whose density is monotonically decreasing, and consequently the distribution of the sample minimum quickly converges to the populational minimum. Since our proposed estimators are semiparametric, we can expect the estimates here to conservatively underestimate the populational minimum.\footnote{An even worst case would be the Pareto distribution, for example, whose density near the populational minimum can be much steeper. We do not analyze this case, but the user of our methods should keep these shortcomings in mind.} For an exponential with rate $\lambda = 1/3$, we obtain the results shown in Fig.~\ref{fig:test-exponential}. For each sample size, a sample was generated from an $\mathit{Exp}(1/3)$ and the estimates $\hat{c}_i$ were calculated; these estimates were then subtracted from the sample, and an exponential distribution was fit by MLE to the modified sample. The log-likelihoods obtained in this process were subtracted from the log-likelihood achieved by $\mathit{Exp}(1/3)$ itself on the original sample, so positive values mean that the estimator did not worsen the likelihood. The subtraction $\hat{l}_1 - \hat{l}_2$ here is related to the ratio of the likelihoods; in fact, taking the exponential of the $\hat{l}_1 - \hat{l}_2$ yields the ratio itself. This experiment has been replicated hundreds of times, and the results are shown in Fig.~\ref{fig:test-exponential} (left), where it can be seen that, on average, the estimators tend to lead to a higher likelihood, which is a good indicative.

High likelihoods are not necessarily good here, due to overfitting. Fig.~\ref{fig:test-exponential} (right) shows the ``distance'' of the estimates from the $5\%$ quantile of the sample minimum distribution for the $\mathit{Exp}(1/3)$; negative values here show an estimate that is below this quantile. If the $5\%$ quantile is deemed as the ideal value, then the estimator $\hat{c}_2$ achieved the best and most steady estimates. All estimators eventually converge to values very near the $5\%$ quantile, meaning that our objectives are indeed being met. We also observed that, for lower values of the rate parameter $\lambda$ (higher variance), the estimators tend to further underestimate the populational minimum, but still converge to the same value. In general, no behaviour that could negatively impact practical scenarios was observed.

The second worst-case scenario considers the Cauchy distribution, which is a heavy-tailed distribution supported on the real line. The fact that its support is infinite, rather than semi-infinite, reflects cases where the experimenter believes a populational minimum exists, but it does not or is much lower than anticipated. Also, being a heavy-tailed distribution, the low quantiles of its sample minimum move relatively fast towards negative infinity. Even so, we argue that the proposed estimators yield good, ``desirable'' results. First because the distribution is heavy-tailed in both sides, so it often yields a large sample variance, which in turn is included in the estimators' equations, so this is factored in. Second, we show that the estimates do not explode to negative infinity; rather than that, it gives estimates that are near the sample minimum to an extent that can be useful to the experimenter that is using positive-supported distributions.

Fig.~\ref{fig:test-cauchy} tries to convey the locations of the estimates by showing the cdf $F(\hat{c} \mid \theta)$ applied at the estimates. $\hat{c}_1$ does not appear here because it is multiplicative and, as discussed in Sec.~\ref{sec:proposed-methods}, only works if the underlying variable is positive. First note on Fig.~\ref{fig:test-cauchy} that the variance is extremely high at low sample sizes (the figure shows only $10\%$ of the standard deviation), which is undesired, but expected. In this sense, only $\hat{c}_4$ had good performance by giving estimates with desirable values of $F(\hat{c} \mid \theta)$ (about $0.03$) even on small samples, although it could be considered a problem that it is too far from $\overbar{\mathscr{m}}$. As discussed in Sec.~\ref{sec:problem-formalization} and illustrated in Fig.~\ref{fig:areas-between-functions}, a low $F(\hat{c} \mid \theta)$ should promote better results in MLE, but can lead to more difficulty in the optimization process. At sample sizes of $50$ and beyond, the variance of the estimates achieve reasonable levels. Again, no anomalies that could hinder practical scenarios were observed.

\begin{figure*}[htb]
    \centering
    \includegraphics[width=0.4\textwidth]{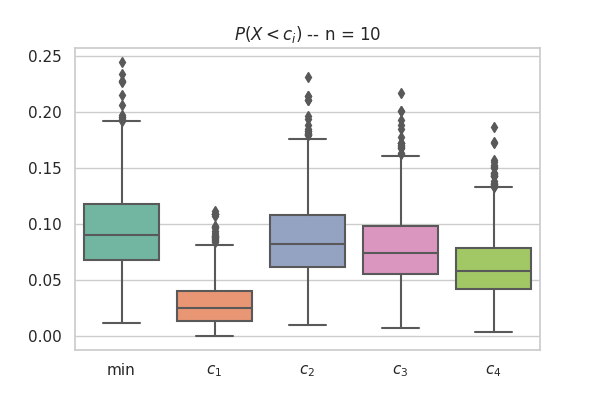}
    \includegraphics[width=0.4\textwidth]{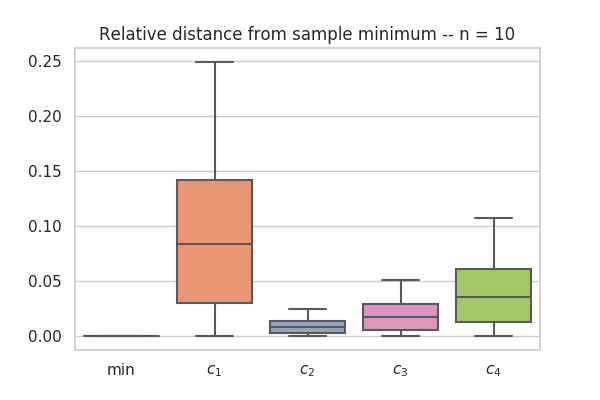}
    \includegraphics[width=0.4\textwidth]{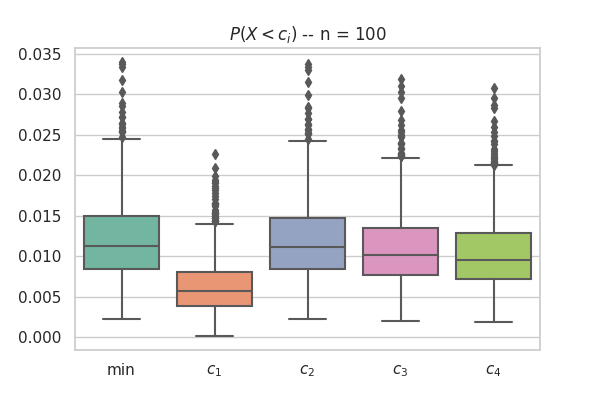}
    \includegraphics[width=0.4\textwidth]{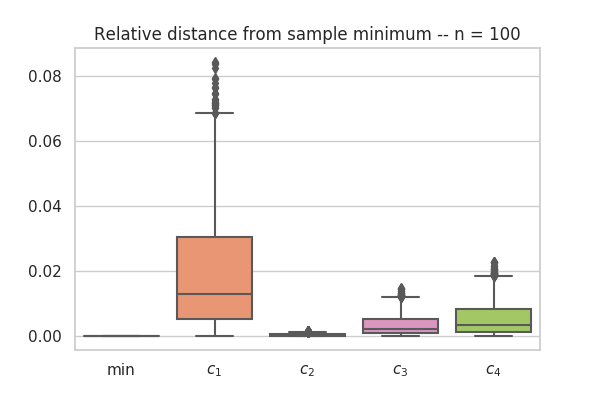}
    \caption{Relative positions of the estimators $\hat{c}_1$, $\hat{c}_2$, $\hat{c}_3$ and $\hat{c}_4$ for many different distributions. Shows the $F(c_i)$ for each estimator (left) and the relative distance from the sample minimum (right). The top portion refers to experiments with samples of size $n = 10$; the bottom refers to $n = 100$.}
    \label{fig:estimators-experiment1}
\end{figure*}

\begin{figure}[htb]
    \centering
    \includegraphics[width=\linewidth]{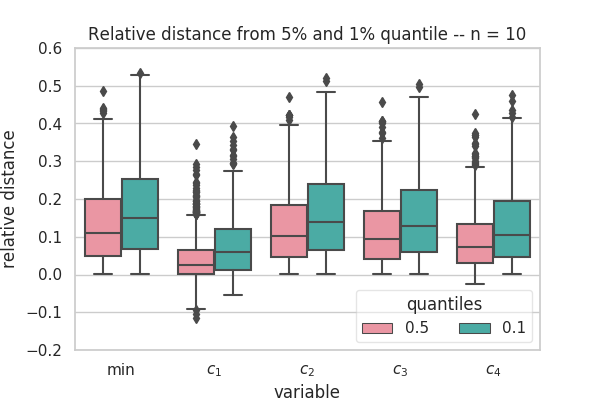}
    \caption{Relative distances from each estimate $c_i$ to the $5\%$ and $1\%$ quantiles of the sample minimum distribution.}
    \label{fig:distance-min-quantiles}
\end{figure}

\medskip
\noindent \textbf{D) Extensive synthetic scenarios}
\smallskip

\noindent We also extensively tested the estimators under distributions of various shapes and scales. Random samples were generated from the Kumaraswamy complementary Weibull geometric (Kw-CWG) distribution, which includes many models as particular cases: gamma, exponential and generalized Weibull distributions, to name a few (see \cite{afify2017new}). A grid of distribution parameters was defined, and for each combination of parameters, 10 sets of $n$ samples were generated and the estimates $\hat{c}_i$ were calculated for each of them. We collected the averaged metrics: i) $F(\hat{c}_i \mid \theta)$ the cdf on the estimator, ii) the relative distance of $\hat{c}_1$ to the sample minimum (normalized by dividing by the populational mean), and iii) the relative signed distance of $\hat{c}_i$ to the $1\%$ and $5\%$ quantiles of the sample minimum. In an attempt to be representative, the grid was defined so that distributions of as various shapes as possible were generated; initially there was a total of $1\,512$ distributions, though some were discarded due to numerical difficulties (generation of \texttt{NaN}s and infinity in the data samples), resulting in $1\,081$ distributions effectively considered.


The experiments show that $\hat{c}_1$ (the multiplicative one, here with $k = 10$) is in general a lot more distant from the sample minimum, and consequently yields a lower cdf $P(X < \hat{c}_1)$, as seen in Fig. \ref{fig:estimators-experiment1}. Estimators $\hat{c}_2$ (involving $1/n$) and $\hat{c}_3$ (based on the iterated logarithm) seem to yield estimates that are too near the actual sample minimum, even for low sample sizes, which is not very desirable. Estimator $\hat{c}_4$ (based on the inequality by Dvoretzky et al., here with $\nu = 0.05$) seems to give a nice balance between these extreme cases. Moreover, the way the distance changes from $n = 10$ to $n = 100$, in the right portion of Fig. \ref{fig:estimators-experiment1}, can be seen as reflecting our expectation that the distance should be larger when we have a small sample, otherwise we have a high risk of $P(X < \hat{c}_i)$ being too high, consequently impairing inference (see Sec.~\ref{sec:problem-formalization}). Experiments with $n = 20$ and $n = 50$ have also been performed, but no particularly different results were observed, relative to the discussion above.

An alternative perspective is obtained if we consider the relative distances from the $5\%$ and $1\%$ quantiles of the sample minimum, shown in Fig. \ref{fig:distance-min-quantiles}. Here we see that most estimators deviate from the these quantiles; if the objective was formulated using these quantiles, only the $\hat{c}_1$ estimator would be reasonable, maybe also $\hat{c}_4$, whose relative distance is mostly kept below $30\%$. We observed that for increasing $n$ all estimates get slightly worse, though $\hat{c}_1$ remains being a fairly reasonable estimate for the $5\%$ quantile. For high $n$, no matter the distance from these quantiles, all estimates will be such that $P(X < \hat{c}_i)$ is very low, so according to the discussed in Sec. \ref{sec:problem-formalization} these worsening relative distances can be disregarded. That is, the sample minimum quantiles might be useful only up to some value of sample size.
\section{Some Notes on Related Work}
\label{sec:related-work}

As discussed in detail in previous sections, our objective is not only to find a low quantile $x_q$, but also ensure it is not too far away from $\overbar{\mathscr{m}}$; and also find it preferably in a non-parametric way. While there is a broad literature in quantiles and their estimation, there does not seem to be related work with the same objectives as ours. This is somewhat understandable because:
1) for practical purposes, subtracting an arbitrary value from the samples is sufficient to workaround the problem of dealing with data with high location and low variance; and 2) the case illustrated in Fig.~\ref{fig:diagram-main-scenario}, where there are multiple datasets to fit a distribution to, does not take place often, so it did not catch enough attention thus far.

With that in mind, the area of quantile estimation is the most related to our work, with some intersection with extreme value theory also. These areas aim at ensuring that a certain random variable will not exceed (or fall below) a certain value, with extreme value theory providing guarantees very close to probability $1$~\cite{haan2007extreme}. This is indeed important for fraud detection \cite{zhang2008detecting}, portfolio optimization \cite{abbasi2013bootstrap} and control of nuclear processes \cite{shockling2015non}, for example, but we do not share the same motivation. Furthermore, they all seek a specific quantile $x_q$, while we are interested in any value within a certain range of quantiles. Despite all these differences, the methods they use have inspired our proposals, so in the following we review the literature in quantile estimation and extreme value theory.

The simplest quantile estimator is $F_n^{-1}(q)$, with $F_n$ being the empirical cumulative distribution function (cdf). Since $F_n$ is a step function, its inverse will lead to a range of possible values for $F_n^{-1}(q)$, and any of them is an estimator for the $q$-quantile $x_q$. This is strengthened by Bahadur's results \cite{serfling1980approximation} that give the convergence in distribution:
\begin{align} \label{eq:bahadur-convergence}
    F_n^{-1}(q) \longrightarrow \mathit{N}\left( x_q, \frac{q (1 - q)}{n f(x_q)} \right)
\end{align}
which holds as long as the derivative $f'$ exists, though this is often relaxed to require only $f$, as done in \cite{dong2017quantile}.
Numerous works use these results to find quantile estimates and confidence intervals thereof \cite{daouia2007nonparametric,koenker2002inference,mu2007power,alexopoulos2019sequest}.
Daouia and Simar \cite{daouia2007nonparametric}, in particular, apply and extend these concepts to the multivariate case, and provide non-parametric results based on existent inequalities on $F_n$. In the context of simulation, the variance of these estimators can be improved by means of Latin hypercube sampling \cite{stein1987large}, which is explored in \cite{kala2019quantile,dong2017quantile,minasny2006conditioned}; Dong and Nakayama \cite{dong2017quantile} combine it with different resampling techniques to propose two estimators with even lower variance. These methods require the underlying cdf to be at least partially given, that is, some mechanism to simulate the underlying variable is needed. Bootstrapping has also been used for variance reduction in quantile estimation, but even so the variance converges very slowly \cite{liu2012convergence}. For our purposes, these methods display a few problems.
First, they require making assumptions on $f$ that allows obtaining the bounds within which $f(x_q \mid \theta)$ has to be, otherwise the variance of the distribution in  Eq.~(\ref{eq:bahadur-convergence}) cannot be determined. Second, even if $f(x_q \mid \theta)$ could be determined, sometimes (particularly when the sample size is large) we will be interested in very low quantiles, which, under mild smoothness assumptions on $f$, would make $f(x_q \mid \theta)$ very close to zero and, consequently, make the variance explode. Third, they are estimates for a fixed quantile, where we would like $q$ to adapt to the sample size, so that the quantile $x_q$ is lower than the sample minimum $\overbar{\mathscr{m}}$.
We could define a rule $q(n)$ that adapts to the sample size $n$, but we decided to stay on the non-parametric path. Also, instead of estimating two values ($q$ and $x_q$), we stick to the idea that it is better to estimate the desired quantity (that is, $\hat{c}$) directly~\cite{vapnik1998statistical}.

Estimation of extreme quantiles, in the context of extreme value theory, relies on a different theoretical ground. Let $Y_{1:n}, Y_{2:n}, \dots, Y_{n:n}$ be the sample order statistics, the distribution of $Y_{[qn]:n}$ can be used to define various estimators for $x_q$, with $[\cdot]$ being any round-to-integer function \cite{mood1950introduction}. The asymptotic distribution of $Y_{1:n}$ and $Y_{n:n}$ is known to belong to the generalized extreme value distribution $\mathit{GEV}(\xi,\allowbreak a,\allowbreak b)$ \cite{haan2007extreme}, and the possible range of parameters can be narrowed down by making assumptions on the tails of the underlying distribution. Taking advantage of this, many extreme quantile estimators arise \cite{valk2018high,drees2003extreme,gardes2018tail,chavez2018extreme}, each requiring different sets of assumptions and yielding estimators with various properties. A fully non-parametric approach, that requires no assumptions, tends not to be possible due to extremely slow convergence (in the worst case) of extreme quantile estimates to the real ones. If the worst case can be expected not to happen, one can rely on kernel density estimation as argued in~\cite{rached2019tail}.

Another popular related area is quantile regression. Although not specifically helpful to this paper's results, it could very well be used to extend our ideas to other scenarios, such as inference on stochastic processes.
In quantile regression, the quantile of interest is from a (usually discrete) stochastic process $(X_n)$. It may be worth noting that in some domains, mainly related to finance, low quantile regression also goes under the name of value at risk estimation \cite{chun2012conditional}. Many approaches begin with an initial estimate $x_q^0$ and update it at every step \cite{alexopoulos2019sequest,he2015approximate}. Some approaches are focused in reducing computational complexity and memory usage, as in \cite{yazidi2017multiplicative,tiwari2019technique,pietrosanu2020advanced}. In \cite{yazidi2017multiplicative} the update rule is as simple as
\begin{align*}
    x_q^{n+1} &= (1 + \lambda q) x_q^n\;\; \text{ if } x_q^n < x^n \\
    x_q^{n+1} &= (1 + \lambda (q-1)) x_q^n\;\; \text{ otherwise }
\end{align*}
and yet achieves incredible performance in some simple synthetic data streams.
Non-incremental approaches include the important results of Koenker and Hallock \cite{koenker2001quantile}, where the ingenious pinball loss function is presented. By means of statistical learning theory \cite{vapnik1998statistical}, Takeuchi et al. \cite{takeuchi2006nonparametric} provide learning guarantees for Koenker and Hallock's method. Recent results include usage of random forests and neural networks \cite{romano2019conformalized}, optimally smoothed pinball loss function \cite{fasiolo2020fast}, and multivariate copula distributions \cite{kraus2017d}. Some of these methods inspired this paper, and we believe our results could also be extended to cover quantile regression in future work.
\section{Conclusion}
\label{sec:conclusion}

We designed the methods presented in Sec.~\ref{sec:proposed-methods} in an attempt to ease the process of performing parameter inference over multiple datasets, a scenario that is illustrated in Fig.~\ref{fig:diagram-main-scenario}. One method was to add the populational minimum as a parameter of the distribution families, and then find it by means of MLE. Another was to iteratively use information obtained during the inference procedure in order to estimate the median of the sample minimum. Both yielded interesting results, although sometimes exhibiting undesirable behavior such as overfitting and larger computational time. Also, it requires some modification of the statistician's usual way to code the MLE process, which might be cumbersome. Despite that, both methods are backed by a more solid theoretical reasoning.

The other proposed methods are arguably simpler to implement, some of which also have theoretical reasoning. They consist of subtracting the sample from certain estimates $\hat{c}$, before performing MLE. In summary, these estimates are:
\begin{align*}
    \hat{c}_1(x_1, \dots, x_n) &= \overbar{\mathscr{m}} \cdot \left( 1 - \frac{\hat{\sigma}}{\hat{\mu} \log_k(n)} \right) \\
    \hat{c}_2(x_1, \dots, x_n) &= \overbar{\mathscr{m}} - \frac{\hat{\sigma}}{n} \\
    \hat{c}_3(x_1, \dots, x_n) &= \overbar{\mathscr{m}} - \hat{\sigma} \cdot \sqrt{\frac{\ln \ln n}{2 n}} \\
    \hat{c}_4(x_1, \dots, x_n) &= \overbar{\mathscr{m}} - \hat{\sigma} \cdot \sqrt{\frac{-\ln(\nu / 2)}{2n}}
\end{align*}
where $\overbar{\mathscr{m}}$ is the sample minimum, $\hat{\mu},\hat{\sigma}$ are the sample mean and variance, $k$ is an arbitrary logarithm basis (we used $10$) and $\nu$ is an arbitrary low probability (we used $0.05$). Of these estimators, only $\hat{c}_2$ displayed occasional unsatisfactory results, and $\hat{c}_1$ is limited to the case where the random variable is supported on $[0, \infty)$ or similar (with slightly different origin).

Based on the experiments shown in Sec.~\ref{sec:4-experiments}, we believe the methods manage reasonably well to achieve the objectives posed initially, and we believe and hope that they are successful in easing the inference tasks of other statisticians and practitioners.

Future work will focus on proving the asymptotic properties of these estimators, as well as extend them to the multivariate cases, also possibly analyzing the particular case of copula models~\cite{nelsen2007introduction}. In order to do that, the the inequalities mentioned in Sec.~\ref{sec:proposed-methods} must be generalized to the multidimensional case, and there is more than one way to do this~\cite{vapnik1998statistical}, which could pose a problem.

\begin{acknowledgements}
We thank prof. Ricardo Marcacini for valuable suggestions, as well as the LaSDPC and BioCom laboratories for the computational and other resources. We also thank CeMEAI (FAPESP grant 2013/07375-0) for providing access to their supercomputer.
\end{acknowledgements}

%
\section*{Conflict of interest}

The authors declare that they have no conflict of interest.

\bibliographystyle{spmpsci}      
\bibliography{references}   

\end{document}